\documentstyle[12pt]{article}
\begin{document}
\phantom {.}
\normalsize
\begin{center}
{\Huge \bf Launching of Non-Dispersive Superluminal Beams}\\[3mm]
\vspace*{1cm}
{\bf    V. S. Barashenkov$^1$ and  W. A. Rodrigues, Jr.$^2$}\\
 \end{center}

\vspace*{3mm}

$^1$  Joint Institute for Nuclear Research, Dubna Moscow Region,\\
\phantom{aaaaa}141980, Russia\\
\phantom{aaaaa}e-mail: barashenkov@lcta30.jinr.dubna.su\\
\phantom{aaa}$^2$ {} Institute of Mathematics,Statistics and Scientific
Computation \\
\phantom{aaaaa}IMECC --- UNICAMP; CP 6065 13081-970, Campinas, SP, Brasil\\
\phantom{aaaaa}e-mail: walrod@ime.unicamp.br

\vspace*{3mm}
\begin{center}
{\bf Abstract}
\end{center}

{\small In this paper we analyze the physical meaning of sub- and
superluminal soliton-like solutions (as the $X$-waves) of the
relativistic wave equations and of some non-trivial solutions of the free
Schr\"odinger equations for which the concepts of phase and group
velocities have a different meaning than in the case of plane wave
solutions. If we accept the strict validity of the principle of
relativity,  such solutions describe objects of two essentially
different nature: carrying energy wave packets and inertia-free properly
phase vibrations. Speeds of the first-type objects can exceed the plane
wave velocity $c_*$ only inside media and is always less then the vacuum
light speed $c$. Particularly, very fast sound pulses with speeds
$c_*<v<c$ have already been launched.  The second-type objects are
incapable of caring energy and information but have superluminal speed.
If we admit the possibility of a breakdowns of Lorentz invariance,
pulses described, for example, by superluminal solutions of the Maxwell
equations can be generated.  Only experiment will give the final
answer.}

\vspace*{3mm}

\vspace*{3mm}

{\bf 1. INTRODUCTION}\\

        As it is known (see, e. g.,~\cite{1} for a review and
references) the hypothesis  of faster-than-light  bodies (tachyons)
contradicts drastically the causality principle if we suppose the
strict validity of the principle relativity. One of the logical
contradictions which exists in the standard formulation of the tachyon
theory is that accepting the validity of the principle of relativity for
faster-than-light bodies  permits us in some situations to  send
informations to the past. E. Recami argued~\cite{2,3} that situations
like that never occurs in reality, however, his arguments are not very
convincing as discussed by many authors (see the review~\cite{1}).  The
second serious difficulty of tachyon theories is the impossibility of a
non-contradictory superluminal generalization of the Lorentz
transformations what is necessary  for any consistent description of the
tachyon kinematics~\cite{4}.

        Meanwhile tachyon-like objects appear in various string
 models, in theories with high-order lagrangians, by supersymmetric
 generalizations. Many authors are of the opinion that this fact is  not
 only a disappointing theoretical failure and  think that an improvement
 of our view of the universe that produces a space-time model compatible
 with superluminal phenomena and free of logical contradictions is
 necessary. Researches in this direction can be found, e.g., in papers~
 \cite{2,6a}.

         There are now many strict mathematical investigations proving
the existence of families of non-dispersive wave packets propagating in
homogeneous media even in vacuum, with arbitrary superluminal speed
$v>c=3.10^{10} cm/s$ (see, e.g., papers~ \cite{6,7,8,9,10,11,13a} and,
especially, a review~\cite{12}) \footnote{Now we don't interest by
priority questions and cite only the papers where one can find the more
detailed bibliography.}.  Such packets correspond to solutions of the
homogeneous wave equations, Klein-Gordon, Dirac, Weyl and Maxwell
 equations which are, dispersion-free, i. e. in contrast to the usual
wave packets made of superpositions of plane waves they don't spread
even in media and, therefore, can be considered as completely
independent space-time localized material objects ("wave torpedos",
purely electromagnetic particles etc.). We call these kind of solutions
"undistorted progressive waves" (UPWs), a name suggested in~\cite{13a,12}
and which seems to the authors to correctly express the essence
of their nature.

   It is important to emphasize that like the plane-wave solutions of
   the relativistic wave equations UPWs have infinite energy. However,
it is possible to exhibit arbitrary $(0\le v<\infty)$ speed solutions of
such equations that have finite energy. Making special superposition of
UPWs, in particular, in some cases using the Rayleigh-Sommerfeld theory
of diffraction, it is possible to obtain finite aperture approximations
(FAA) to a given solution of the relativistic wave equations that have
finite energy~\cite{13a,12}. One can verify also that such finite energy
solutions (subluminal, luminal or superluminal)  are quasi-undistorted
progressive waves (QUPWs). For the QUPWs solutions of the Maxwell
equations it can be proved that they decay into solutions travelling
with the usual light speed $c$. In this sense QUPWs looks like instable
particles.

   The existence of QUPWs-type solutions for the case of sound has been
proved by  J.-Y. Lu and J. E. Greenleaf~\cite{11}. The existence of
QUPWs satisfying eq. (1.1) and traveling with speeds either $v<c_*$ or
$v>c_*$ where  $c_*$, called speed of sound, is a characteristic of the
properties of the medium, e.g., the temperature, Young modulus, etc, is
proved also experimentally~\cite{12,13} \footnote{In the experiments
the so-called X-wave packet travels only with speed 0.2441(8)\% greater
than $c_*$ and skeptics can have doubts. A new experiments with QUPWs
having speeds $V\gg c_*$ should be done. Such "superfast waves",
especially in water, are of great practical importance.}.

	It is very important for all considerations that follows to
take into account the following. Usually the velocity of propagation of
energy of a wave satisfying the equation

$$ \frac{1}{c_*^2} \frac{\partial^2\Phi}{\partial t^2}-
\nabla^2\Phi=0,  \eqno (1.1) $$
where $\Phi(t,{\bf x})$ is the pressure at the point $\bf x$ at the time $t$
is defined as

  $${\bf v_e}={\bf S}/u.   \eqno(1.2)    $$
Here $\bf S $ is the momentum flux and $u$ is the energy density given
by

  $$ {\bf S}=\nabla\Phi\partial\Phi/\partial t,  \quad
  u=(1/2)[(\nabla\Phi)^2+(1/c_*^2)(\partial\Phi/\partial t)^2]. \eqno(1.3) $$
 From these expressions it follows that
      $$      v_e \le c_*             \eqno(1.4) $$

      One can easily prove that for plane waves $v_e=c_*$ indeed.
      However, the acoustic experiments described in the papers
\cite{12,13} show that the FAA to the limited band sound X-wave travels with
a speed $v>c_*$.  This speed is {\it also} the speed of the propagation
of the energy carried by the wave, since the hydrophone located at a
distance $d$ from the transducer is activated by the energy carried by
the wave at a time $t=d/v$ after the launching of the wave at a moment
$t=0$. It means that the definition (1.2) is devoid of sense. For the
case of the FAA to the limited band sound X-wave the speed $v_e$ is a
complicated function of $({\bf x},t)$, so we must be careful when
discussing the velocity of a propagation of energy of UPWs and QUPWs
solutions, particularly, of Maxwell equations.

   From the viewpoint of current quantum theory a sound wave is composed of
phonons  and excitations that travel in the  medium with speed $c_*$.
The problem of understanding the acoustic waves travelling with speeds
$v<c_*$ or $v>c_*$ using quantum field theory will be  analyzed in another
paper where we  show that they correspond to a new kind of boson
condensate.

         Another important remark is necessary to be emphasized here.
         The superluminal UPWs or QUPWs solutions of the free Maxwell
equations share with the above mentioned classical tachyons only the
property that both travel with speed $v>c$. The analogy ends here
because any classical tachyon  is a  material object (an elementary
particle, a macroscopic body) whose Lorentz change of the longitudinal
dimension in the laboratory frame is defined by the cut of the tachyon
world tube by a plane $t=\mbox{constant}$. Since tachyons are always
moving we can write
 $$ L(v)/L(v')=(v^2/c^2-1)^{1/2}/(v'^2/c^2-1)^{1/2}.  \eqno(1.5)  $$
This means that the tachyon mass and energy distributions in the
laboratory frame must be ellipsoidal also, but in contrast to the case
of subluminal bodies instead of a contraction the mass and energy
distributions will show a dilation which increases with the tachyon
speed $v$ \cite{14}. It differs essentially from the X-like form of some
UPWs or QUPWs solutions of the wave equation.  The conclusion about
X-like  shape of the moving tachyons obtained by E. Recami~\cite{2} is a
consequence of the formal Lorentz transformation to the laboratory frame
from a superluminal reference frame. However, as mentioned above, such a
transformation is contradictory. So, the coincidence of the tachyon
shape predicted  by E.  Recami with the X-wave solutions of the Maxwell
equations should be considered as accidental and doesn't prove the
breakdown of the well known relativistic shape change law for the
tachyons.

    After these remarks we complete the introduction by saying that the
    main purpose of our paper is to analyze the physical meaning of the
various wave velocities that appear (i) in some extraordinary solutions
of the Schr\"odinger equation (Section 2), (ii) in  the case of  UPWs
$v>c_*$ solutions of the wave equation for sound waves (Section 3) and
(iii) in UPWs $v>c$ solutions of the free Maxwell equations (Section 4)
where, in particular,  we analyze the energy velocity paradox quoted by
W. Band~\cite{8}.

   Obviously, with the analysis of the superluminal electromagnetic UPWs
   in Section 4 we are not proving that it is sure that such waves
can be launched in physical space, and, of course,  if we believe in the
strict validity of the principle of relativity which is one of the main
dogmas of current physics, to launch a "wave torpedo" with $v>c$ is
impossible.  Nevertheless the detailed computer simulation to the finite
aperture approximation to the superluminal electromagnetic waves
presented in the paper~\cite{12} suggests that it is worth to try the
experiment. Sure, if the FAA to a given superluminal electromagnetic
wave can be launched, we will see a breakdown of the principle
relativity as it is known today, thus determining the limits of validity
of the special theory of relativity.\\[3mm]

 {\bf 2. NON-DISPERSIVE SOLUTIONS OF SCHR\"ODINGER EQUATION}\\

    In dealing with waves we have several velocities involved, e. g.,
phase and group velocities, velocity of transport of energy, signal speed,
front velocity etc... Of course, the meaning of all these quantities
depends on the particular theory to which the wave motion is associated.
In order to reveal their relations more clear let us take the case of
non-relativistic quantum mechanics where a free particle has associated to
a wave function $\Psi$ satisfying the equation
  $$ i\hbar \frac{\partial\Psi}{\partial t}+
      \frac{\hbar^2}{2m}\nabla^2\Psi=0.   \eqno(2.1)   $$
Now, if the particle is moving with the kinetic energy $E=mv^2/2$ and the
momentum ${\bf p}=mv{\bf z}$, then the associated wave function in the
solution of eq. (2.1) given by the plane wave
$$ \Psi=Ae^{i(\omega t-kz)}      \eqno(2.2)    $$
with A is a constant and
$$ E=\hbar\omega=mv^2/2, \quad p=\hbar k=mv   \eqno(2.3)   $$

          The function $\Psi$ is simultaneously eigenfunction of  energy
 and momentum operators:
 $$ {\hat E}\Psi \equiv (i\hbar\partial/\partial t)\Psi=\hbar\omega\Psi,\quad
 {\hat p}\equiv\Psi=(-i\hbar\partial/\partial z)\Psi=
 \hbar k\Psi. \eqno(2.4) $$
The connection between $\omega$ and $k$ in order for eq. (2.2) to be a
 solution of eq. (2.1) gives the dispersion relation
 $$  \omega=\hbar k^2/2m.   \eqno(2.5) $$

      The phase and group velocities associated with $\Psi$ are
$$ v_{ph}=v/2, \quad v_g=d\omega/dk=v  \eqno(2.6)  $$
and we see that for plane waves the speed of the particle is equal to $v_g$.

   Now, as it well known, the function $\Psi$ given by eq. (2.2) is not
   an element of the Hilbert space $(H)$ of wave functions. A probability
   wave $\Psi(z,t) \in H$ describing a particle moving
with speed ${\bf v}=v{\bf z}$ is given by the wave packet
$$ \Psi(z,t)=A\int dk B(k)e^{i(kz-\omega t)},  \eqno(2.7)  $$
where $B(k)$ is a weight function centered in $k_o$ and decaying
rapidly outside the interval $k_o-\Delta k< k<k_o+\Delta k$.
For $\Psi $ given by eq.(2.7) the group velocity is defined by
$$ v_g=(d\omega/dk)_{k_o}. \eqno(2.8)  $$
The  function $\Psi $ is than interpreted as associated with a particle
moving with expectation kinetic energy and momentum
 $$E_o=mv^2/2=\langle \Psi| {\hat E} |\Psi\rangle, \quad
 p_o=mv_o=\langle \Psi| {\hat p_z} |\Psi\rangle.     \eqno(2.9) $$

    The important point to be emphasized here is that $\Psi $ is a
    {\it spreading}  wave packet, and the question naturally arises
    here: are there any non-spreading wave packets which are solutions
    of Schr\"odinger equation? If the answer is positive, which is
    the meaning of such solutions?

     If we solve eq. (2.1) in cylindrical co-ordinates for a wave moving
along the $z$-axis, we immediately get a family of solutions
$$ \Psi_{J_n}=A_n e^{in\theta}J_n(\alpha \rho)e^{i(kz-\omega t)},
\eqno(2.10) $$
where $A_n$ is a constant, $\rho=(x^2+y^2)^{1/2}$ and $\alpha$ is the
so-called separation constant what means that $\alpha$ is not a function of
$({\bf x},t)$ but, of course, it may be a general function of $k, \omega$
and other parameters.

      In what follows let us consider for simplicity the  solution with
 $n=0$.

 The dispersion relation
     $$ \omega =\frac{\hbar}{2m}(k^2+\alpha^2)   \eqno(2.11) $$
 must hold in order for  the function $\Psi_{J_o}$  to be a solution of
 eq. (2.1). To interpret the physical meaning of this relation we recall that
 $\Psi_{J_o}$ is a simultaneous eigenfunction of ${\hat E}$ and  ${\hat p}$.
  Indeed, we have
$${\hat E}\Psi_{J_o}=\hbar \omega\Psi_{J_o}=E\Psi_{J_o}, \quad
 {\hat p}\Psi_{J_o}=\hbar k\Psi_{J_o}=p\Psi_{J_o}.    \eqno(2.12) $$
Then, we may write
$$ E=\frac{\hbar^2}{2m}(k^2+\alpha^2)=
\frac{\hbar^2}{2m}+mc^2, \quad p=\hbar k \eqno(2.13) $$
where we put
$$ m=\hbar\alpha/\sqrt{2}c  \eqno(2.14) $$
interpreted as the mass of the particle. We have then, as usual, the particle,
phase and group velocities
$$v=p/m=\hbar k/m=\sqrt{2}ck/\alpha, \quad v_{ph}=\omega/k, \quad
v_g=d\omega/dk=v \eqno(2.15) $$

    Let us now introduce, following  J.-Y. Lu and J. F. Greenleaf~\cite{13},
the axicon angle $\eta$ by
$$ k=\omega\xi\cos{\eta}, \quad \alpha=\omega\xi\sin{\eta}  \eqno(2.16)  $$
where $\xi$ is determined by the dispersion relation (2.11) as
$$\xi=(2m/\hbar\omega)^{1/2}=\sqrt{2}c^{-1}\sin\eta. \eqno(2.17)  $$

               Taking into account now the relation between $\alpha$
and $m$, given by the eq. (2.14) we have
    $$ \omega=m c^2/\hbar \sin^2\eta  \eqno (2.18)    $$
Then we have for the velocities in the eq. (2.15:
    $$ v=\sqrt{2}c\cot{\eta} \eqno(2.19) $$
    $$v_{ph}=v_g=\sqrt{2}c/\sin{2\eta}=v/2\cos^2\eta   \eqno(2.20)   $$
In contrast to the plane wave solution here $v_{ph}=v_g$, but in general
$v_g\ne v$. For all this, as the Schr\"odinger equation is a non-relativistic,

       $$ v/c=\sqrt{2}\cot{\eta}\ll1, \eqno(2.21)   $$
i. e. the axicon angle has to be close to $\pi/2$, and, respectively,
             $$ v\simeq\sqrt{2}c\cos{\eta}\ll c  \eqno(2.22)   $$
$$ v_{ph}=v_g\simeq \sqrt{2}c/\cos{\eta}. \eqno(2.23)   $$

    The interpretation of the velocities are now clear. If we want that $\Psi_{J_o}$
describes a free particle moving with the speed $v$, then we cannot attach a
physical meaning to $v_{ph}$ or $v_g$, i. e. in this case the transport of energy
will not be given by $v_g$.

          We can now  construct a wave packet
 $$\Psi_n=A_ne^{in\theta}\int d\omega B(\omega)
 J_n(\omega\sqrt{2}c_{-1}\rho\sin^2\eta)
 e^{i\omega(2zv^{-1}\cos^2\eta-t)}.  \eqno(2.24)   $$
which is, obviously, dispersive solution of eq. (2.1) if we consider
the mass (2.14) as a constant, because in this case the axicon angle $\eta$
and, respectively, the velocities (2.19), (2.20) are frequency dependent.
In this connection it should be noted that the results presented in the
paper~\cite{17} claiming the existence of X-wave solutions for the
Schr\"odinger equation are seen unfortunately to be incorrect for analogous
reasons.

   In a paper~\cite{18} A. Barut tried to construct non-dispersive solutions of the
Schr\"odinger equation using spherical symmetry. If we suppose that there is
a stationary solution of eq.(2.1) of the form
$$\Psi({\bf x},t)=e^{-imc^2t/2\hbar}f({\bf x})      \eqno(2.25)  $$
then putting $\Psi$ into  eq. (2.1) we see that the function $f(\bf x)$
satisfies the Helmholtz differential equation
$$\nabla^2f({\bf x})+\alpha^2f({\bf x})=0   \eqno(2.26)   $$
with $\alpha=mc/\hbar$.

       A simple solution of this equation in spherical coordinates is
  $\sin \alpha r/r$ where $r=(x^2+y^2+z^2)^{1/2}$. Then, really a
non-dispersive packet exist which is stationary in the laboratory frame and
is given by
$$\Psi({\bf x},t)=\frac{\sin\alpha r}{r}e^{-imc^2t/2\hbar}.  \eqno(2.27)
$$

   It is clear that $\Psi({\bf x},t)$ given by eq. (2.27) is not an
   eigenfunction of the momentum operator, but, nevertheless, it is an
eigenfunction of the energy operator.

  A. Barut claims to have constructed a non-dispersive  wave packet solution
 of the Schr\"odinger equation by first writing
$$ \Psi({\bf x},t)=\psi({\bf x},t)e^{-imc^2t/2\hbar}  \eqno(2.28) $$
where the function $\psi({\bf x},t)$ satisfies the equation
$$ 
i\hbar\frac{\partial\psi}{\partial t}+
\frac{\hbar^2}{2m}\nabla^2\psi + \frac{mc^2}{2} \psi 
=0.  \eqno(2.29) 
$$
Then the author of the paper~\cite{18} used a ansatz
$$ \psi({\bf x},t)=f(\xi)\exp{[-i\hbar^{-1}mv(z-vt/2)]}  \eqno(2.30) $$
with $\xi=[x^2+y^2+(z-vt)^2]^{1/2}$. Substituting this expression into
eq. (2.29) gives
$$ 
i\hbar\frac{\partial}{\partial t}f(\xi)+
\left[\frac{\hbar^2}{2m}\nabla^2f(\xi)+
\frac{mc^2}{2}f(\xi)\right]=0.   \eqno(2.31)   
$$

   As the next step A. Barut proposed to put
   $$ \nabla^2f(\xi)+(mc/\hbar)^2f(\xi)=0. \eqno(2.32)   $$
However,to be satisfied it is necessary that
$$ \partial f(\xi)/\partial t=0.  \eqno(2.33)   $$
This equation has a solution only when $v=0$ which, particularly,
produces as a possible solution our stationary wave packet given
by eq. (2.27).

We can think that an $X$-wave solution of Schr\"odinger equation can be
constructed by relaxing the condition that $\alpha$ in eq. (2.13) is a
constant. Unfortunately even in this case a simple calculation shows
that there is no such a solution. Nevertheless, it is worth to try to
construct ion beams with the wave function like that given in eq.~(2.10)
in an experiment analogous to a light experiment  by Durnin \cite{7} but
using instead of an optical lens a magnetic one. 

{\bf 3. UPWs SOLUTIONS OF THE WAVE EQUATION}\\ \nopagebreak

     The solution of the homogenous wave equation in cylindrical
     coordinates has the same form (2.10) but with the different
 dispersion relation
 $$   \omega^2=c_*^2(k^2+\alpha^2).
 \eqno(3.1)
$$

     Consider a quantum mechanical meaning for these expressions. If we
write $E=\hbar\omega$ and $p_z=c_*c^{-1}\hbar k$ then
$$ E^2/c^2-p_z^2=m^2, \quad m=\hbar\alpha c_*/c. \eqno(3.2)  $$
Putting now
$$k=c_*\omega\cos\eta,\quad \alpha=c_*\omega\sin\eta \eqno(3.3)  $$
we can define the velocities
$$ v=p_zc^2/E=c\cos\eta,   \eqno{(3.4)} $$
$$ v_{ph}=v_g=\omega/k=c_*/cos\eta.   \eqno(3.5) $$

      We see that the massless wave equation has solutions propagating with phase
and group velocities (3.5), and $v$  would be the velocity of a
excitation-like particle with the mass depending on the frequency:

$$ m=\hbar c_*c^{-1}\omega \sin\eta. \eqno(3.6) $$

     Using relations (2.10) and (3.5), one can build the packet
$$
\Phi_{Xn}=A_n e^{in\theta}\int d\omega B(\omega)
J_n(c_*^{-1}\omega \rho\sin\eta)
e^{i\omega(c_*^{-1}\cos\eta-t)}
\eqno(3.7)
$$
which moves rigidly, without  any distortion:
$$
\Phi_{X_n} (x,y,z,t)=\Phi_{X_n} (x,y,z-(c_*/\cos{\eta})t,0).
\eqno(3.8)
$$

 We remark that the packet $\Phi_{Xn}$, if interpreted, for example,
 as a classical sound  wave, has infinite energy. However,
 a finite approximation to $\Phi_{Xn}$ with  an appropriate function
 $B(\omega)$  has been seen to travel with the  speed
 $c_*/\cos{\eta}$~\cite{12}. Thus, for this case, as said in
 Section 1, this is the velocity of propagation of the wave energy ---
 a non-trivial fact showing again that the interpretation of the velocities
 associated to a wave depends on the theory, that the wave is supposed
 to describe.

   From the quantum-mechanical point of view sound is composed of phonons.
 $\Psi_{Xn}$ is a kind of field configuration defining a new kind of boson
 condensate. This will be studied later.\\[3mm]

{\bf 4. SUPERLUMINAL WAVE PACKETS}\\

        If $c_*=c$ the homogenous wave equations have superluminal
solutions with $v_{ph}=v_g>c$ (see eq.(3.5)). As shown in~\cite{13a,12} such
solutions exist also for massive Dirac and Klein-Gordon particles.  If
we strictly believe in the presently known relativistic physics these
superluminal packets, of course, cannot be generated. What, however, is
the physical meaning of such faster-than-light solutions in this case?

      Solutions of this kind describe inertia-free processes like, for
example, a neon advertisement string where each letter flashes
independently of the preceding one. Neither information nor energy is
transferred in these processes and the problem concerning the
velocity of energy transport simply doesn't exist here.

      A superluminal electromagnetic beam can be launched in physical
space with a boundary, like in W. Band's gedanken experiment \cite{8}
where a charged cylinder with an appropriate charge density is used.
Band's solution with $v_g>c$ describes an inertia-free process if we
{\it change} the charge of each tiny cylinder segment ("switch" it up)
quite independently according to Band's solution. One should say that
the situation is here quite clear, again there is no transfer of
information and energy, and Band's problem concerning the  ratio $|{\bf
S}|/u <1$ (of Poynting vector over energy density) is trivially no
problem at all.

  As we said already in the Introduction,  the ``mathematical
  experiments'' done in~\cite{12} for the electromagnetic X-waves
solutions of free Maxwell equations seem, however, to indicate that
eventually these waves can be generated with appropriate antennas. Of
 course,  if this really can be done we will find a violation of the
 principle of relativity.\\[3mm]

 {\bf 5. CONCLUSIONS}\\

   We see that both the sub- and superluminal non dispersive soliton-like
solutions of the homogeneous linear relativistic equations describe
physical situations that can be realized in practice. For superluminal
solutions there is surely a real transfer of energy and information,
superluminal wave packets describe inertia-free processes without any
transfer. Nevertheless it remains to verify if {\it real}, i. e. energy
transferring, faster-than-light electromagnetic pulses suggested by
mathematical simulation can be launched. The properties of UPWs
 are so extraordinary that only experiment can decide, and if such
 beams exist then we have a first case of a breakdown of the current
form of the theory of relativity.\\[3mm]

{\bf  Acknowledgments}\\

One of the authors (V. S. B.) is grateful to the Direction of the
Institute of Mathematics, Statistics and Scientific Computation of
Campinas State University for hospitality and financial support. The
authors are grateful to Dr. E. Capelas de Oliveira, J. E. Maiorino and
 M. Z. Yur'iev for useful discussions. \\

\end{document}